\documentclass[aps,prl,twocolumn,superscriptaddress]{revtex4-1}
\usepackage{blindtext}
\usepackage{bm}
\usepackage{amsmath,amssymb,amsfonts,latexsym,fancyhdr,graphicx,epstopdf,times,txfonts}

\usepackage[pdftex,colorlinks=true,allcolors=blue]{hyperref}
\usepackage[english]{babel}
\usepackage{graphicx}
\usepackage {subfigure,overpic}

\usepackage{xcolor}

\usepackage{verbatim}
\usepackage{lipsum}

\def\bra#1{\mathinner{\langle{#1}|}}
\def\ket#1{\mathinner{|{#1}\rangle}}

\definecolor{greengreen}{HTML}{2F9B31}

\begin{document}
\title{Bose-Hubbard lattice as a controllable environment for open quantum systems}
\author{Francesco Cosco}
\affiliation{Turku Centre for Quantum Physics, Department of Physics and Astronomy, University of Turku, FI-20014 Turun yliopisto, Finland}
\author{Massimo Borrelli}
\affiliation{Turku Centre for Quantum Physics, Department of Physics and Astronomy, University of Turku, FI-20014 Turun yliopisto, Finland}
\author{Juan Jos\'{e} Mendoza-Arenas}
\affiliation{ Departamento de F\' isica, Universidad de los Andes, A.A. 4976,
Bogot\'{a} D. C., Colombia}
\affiliation{Clarendon Laboratory, University of Oxford, Parks Road, Oxford OX1 3PU, United Kingdom}
\author{Francesco Plastina}
\affiliation{Dipartimento di Fisica, Universit\`a della Calabria, 87036, Arcavata di Rende (CS), Italy}
\affiliation{INFN - Gruppo Collegato di Cosenza, Cosenza, Italy}
\author{Dieter Jaksch}
\affiliation{Clarendon Laboratory, University of Oxford, Parks Road, Oxford OX1 3PU, United Kingdom}
\affiliation{Centre for Quantum Technologies, National University of Singapore, 3 Science Drive 2, Singapore 117543}
\author{Sabrina Maniscalco}
\affiliation{Turku Centre for Quantum Physics, Department of Physics and Astronomy, University of Turku, FI-20014 Turun yliopisto, Finland}
\affiliation{Center for Quantum Engineering, Department of Applied Physics, Aalto University School of Science, P.O. Box 11000, FIN-00076 Aalto, Finland}

\selectlanguage{english}

\begin{abstract}
We investigate the open dynamics of an atomic impurity embedded in
a one-dimensional Bose-Hubbard lattice. We derive the reduced
evolution equation for the impurity and show that the Bose-Hubbard
lattice behaves as a tunable engineered environment allowing to
simulate both Markovian and non-Markovian dynamics in a controlled
and experimentally realisable way. 
We demonstrate that the presence  or absence of memory effects is
a signature of the nature of the excitations induced by the
impurity, being delocalized or localized in the two limiting cases
of superfluid and Mott insulator, respectively. Furthermore, our
findings show how the excitations supported in the two phases can
be characterized as information carriers.
\end{abstract}

\maketitle
{\itshape Introduction -} Several condensed matter models have
been  recently investigated from an open system viewpoint.
Typically, one imagines  embedding an  impurity in a much
larger many-body system that acts as its environment. The
underlying idea is that, by monitoring dynamical properties of
such a probe-impurity, one can access information about the
many-body system indirectly and, ideally, with very little
disturbance. Several features of a many-body system can be
thoroughly investigated within this framework, from
single-particle excitation
spectra to many-body correlations \cite{bruderer,kantian,mitchinson,elliot,streif2016}. 

In a complementary fashion, one can shift the perspective and
study properties of the impurity dynamics itself. In this respect,
an interesting question to ask is whether the open dynamics
induced by a many-body system is memory-less or not. Generally,
 the answer  depends on the specific system at
hand, as well as on those controllable parameters that make
tunable the many-body environment. Open system dynamics induced in
many-body experimental platforms have been studied and simulated
mainly in the Markovian, or memory-free, regime
\cite{sabrina1,blatt1,blatt2,weimer}, modelled by a Lindblad
master equation \cite{reviewblp}. In a number of physical
scenarios, however, a Markovian description of the dynamics is
inadequate \cite{jelezko2004,paladino2014,vats1998}. Furthermore,
memory effects may be beneficial for certain quantum-enhanced
protocols, such as superdense coding \cite{antti1,biheng},
teleportation \cite{laine2014}, and quantum key-distribution \cite{vasile2011}, and
they play a key role as well in quantum thermodynamics
\cite{bogna} and measurement theory \cite {karpat2015}. Their
quantification in terms of information backflow has led to the
introduction of a number of non-Markovianity measures or
witnesses, based on the dynamical properties of different quantum
information quantifiers\cite{wolf,blp,rhp,luo,lorenzo1,bogna2}.
Depending on both the system and the quantum protocol of interest,
one may choose the appropriate non-Markovianity measure and
investigate if
and in which way memory effects lead to optimised performance.
In light of this renewed interest in non-Markovian dynamics, a few
experiments in the quantum optical domain have been performed to
implement quantum simulators of simple non-Markovian models
\cite{liu,li,cialdi,tang,chiuri,jin,orieux,bernardes, cialdi2017}.
At the same time a number of theoretical studies have demonstrated
that many-body environments may induce memory effects in the
dynamics of an interacting impurity
\cite{apollaro,lorenzo3,pinja1,sindona,cetina1,cetina2}.  

In this Letter, we focus on the Bose-Hubbard model, which is efficiently
implemented using cold atoms in optical lattices
\cite{dieter1,blochrev,cazalilla}. In particular, the occurrence
of a critical point has been demonstrated, which separates a
superfluid quasi-condensate from a Mott insulator phase
\cite{greiner}. We consider the one-dimensional model, whose phase
diagram has been extensively investigated \cite{kuhn}. In contrast
to previous works on spin system and ion crystals
\cite{pinja2,massimo}, our results show for the first time the
existence of a direct connection between the presence or absence
of memory effects in the induced open system dynamics and the
nature of the environmental excitations, giving rise to either
delocalized or localized density fluctuations in the two phases,
respectively. By using both analytical and numerical tools, we
evaluate the amount of information backflow and its dependence
upon the parameters of the Bose-Hubbard model. In particular, we
adopt two different analytic approaches in the superfluid and deep
Mott regimes, respectively, which allowed us to identify the
single particle excitations in the two phases and to obtain their
contributions to information flow and memory effects. Then we use
numerical t-DMRG to interpolate between the two extremes and
analyse non-Markovianity close to the transition point. Besides
directly linking memory effects to the physical properties of the
information carriers, this analysis offers the possibility of
characterizing the mobility of the excitations from a quantum
information perspective. 

In a related work \cite{lorenzo2},
memory effects induced by Anderson localisation were studied for a
two-level atom coupled to a one-dimensional array of disordered
cavities, and non-Markovianity was shown to increase with the
disorder strength. Taken together with these observations, our
results point towards a possible universal connection between
non-Markovianity and localization in complex many-body
environments.

{\itshape Impurity dephasing in the Bose-Hubbard lattice -} The
dynamics of a cold  gas of bosonic atoms trapped in a
one-dimensional optical lattice and confined to the lowest Bloch
band is governed by the Bose-Hubbard Hamiltonian
\cite{dieter1,blochrev2}
\begin{equation}
\hat{H}_{BH}= -J \sum_{i} (\hat a^\dagger_{i} \hat a_{i+1}+\hat a^\dagger_{i+1} \hat a_{i}) + \frac {U}{2} \sum_i  \hat n_i (\hat n_i -1),
\label{bh}
\end{equation}
where $J$ is the hopping parameter, $U$ is the localising on-site
interaction, 
$\hat{a}_{i}, \hat{a}_{i}^{\dagger}$ are the standard boson
annihilation and creation operators and
$\hat{n}_{i}=\hat{a}_{i}^{\dagger}\hat{a}_{i}$. In the limits
$J=0$ and $U=0$, the above Hamiltonian is exactly solvable, with
the ground states describing either a Mott insulator, with a
uniform  occupation number per site $\bar{n}$, or a
quasi-condensate, with a macro-occupancy of the lowest
momentum-state, respectively. The intermediate regime is
analytically intractable but approximate models can be employed in
the $J\gg U$ and $U\gg J$ regimes
\cite{bhbog,huber2007,barmettler2012}. 

A single atomic impurity
can be used to investigate features of the lattice trapped
gas \cite{recati,cirone,cosco2017}. We assume that the impurity is
harmonically confined in three dimensions, frozen
in its motional ground state, whose wavefunction is a Gaussian
centered at a specific site (say $i=0$) of the lattice. This
setting can be practically implemented using selective optical
potentials that are able to confine different internal states or
different atomic species independently
\cite{ospe,catani,kay,bloch3,spethmann,catani2}. The impurity and
the surrounding gas are coupled via a density-density interaction,
whose strength depends on the internal state of the impurity, and
which couples it to the local number operator $\hat{n}_0$
\cite{bruderer,johnson}. We assume that only the two lowest
internal levels of the impurity ($\ket{e}, \ket{g}$) are relevant
to the dynamics and that $\ket{g}$ does not couple to the gas.
This can be achieved by a Feshbach resonance
\cite{ospe,catani,kay,bloch3,spethmann,catani2,johnson}. The total
Hamiltonian reads
$\hat{H}=\frac{\omega_{0}}{2}\hat{\sigma}_{z}+\hat{H}_{BH}+U_{e}|e\rangle\langle e|\otimes\hat{n}_0$,
where $U_{e}$ is an effective coupling constant accounting for the
spatial overlap between the impurity and the gas wave functions.
We will consider a repulsive interaction, $U_e>0$.

Starting with the impurity initially in state $\ket g$, we assume
it to be transferred into an equal superposition of $\ket g$ and
$\ket e$ by a $\pi/2$ pulse. The ensuing evolution generated by
$\hat H$ will not lead to population transfer. It will, however,
cause dephasing, giving rise to a decay of the impurity coherence.
The central physical quantity of interest in this work will be the
decoherence function, which we will use to characterize the
impurity dynamics. It can be measured in a Ramsey-type experiment
by applying a second $\pi/2$ pulse, after a variable delay $t$,
that maps the relative phase between  $\ket g$ and $\ket e$ into a
measurable population imbalance \cite{Batalhao}. The
environment-induced dynamics of the impurity's internal levels can
be either memory-less (Markovian), with a monotonic loss of the
initial coherence, or with memory (non-Markovian), with some
temporary and partial recovery. Starting from the microscopic
Hamiltonian $\hat H$, by means of standard time convolutionless
projection operator technique \cite{breuerbook}, one can derive a
perturbative master equation for the reduced impurity state,
\begin{equation}
\frac{d \rho}{dt}=-\frac{i\bar \omega_0}{2} [\hat{\sigma}_z, \rho] + \gamma(t) (\hat{\sigma}_z \rho \hat{\sigma}_z - \rho),
\label{mastereq}
\end{equation}
which is well known as a dephasing master equation (with a
renormalized transition frequency $\bar \omega_0$). The
time-dependent transition rate $\gamma (t)$ is directly connected
to the density fluctuations of the gas
\begin{equation}
\gamma (t)=U_e^2 \text {Re} \int_0^{t} \, dt' \langle \hat {n}_0(t') \hat{ n}_0(0) \rangle.
\end{equation}
Through this relation, the properties of the Bose-Hubbard gas are
dynamically mapped into the impurity time evolution, which comes
out to be Markovian if and only if the decay rate is positive at
all times, implying no memory-effects. Non-Markovian dynamics
occurs, instead, if at least one time interval exists for which
$\gamma(t) <0$, causing information backflow \cite{blp}. This is
directly linked to the time behaviour of the Loschmidt echo,
defined as $L(t) \equiv|\bra{\phi(t)}\phi'(t)\rangle|^{2},$ with
$\ket {\phi(t)}$ being  the ground state $\ket{\phi(0)}$ of $\hat H_{BH}$, chosen
as the initial state, evolved according to $\hat H_{BH}$ only,
while $\ket{\phi'(t)}$ is the time-evolved state in the presence
of the impurity. It turns out that
$|\rho_{eg}(t)|/|\rho_{eg}(0)|=|\bra {\phi(0)} e^{i \hat H_g t}
e^{-i \hat H_e t} \ket {\phi(0)}|=\sqrt{L(t)}$, where $\hat H_{k}=
\bra {k} \hat H \ket {k}$, with $k=g,e$. Therefore,
$\sqrt{L(t)}=e^{\Gamma(t)}$, where $\Gamma(t) =- \int_0^t dt'
\gamma (t')$, is uniquely determined by the dephasing coefficient of
the impurity.

In order to investigate the Markovian or non-Markovian character
of the dynamics induced by the engineered quantum environment we
use the trace-distance measure (BLP measure) quantifying
information flow in terms of distinguishability between two
arbitrary quantum states of the probe \cite{blp}. It is worth
stressing that, for the open quantum system investigated here, all
non-Markovianity measures agree in discerning Markovian from
non-Markovian dynamics \cite{reviewblp,competition}. For our case, the BLP non-Markovianity measure $\mathcal{N}$ can
be explicitly computed via the Loschmidt echo \cite{antti2,pinja2}
\begin{equation}
\mathcal{N}=\sum_{n}\sqrt{L(t_{n+1})}-\sqrt{L(t_{n})},
\label{nmopt}
\end{equation}
where the sum is taken over the set of time intervals
$[t_{n},t_{n+1}]$ at which the  echo increases. During these
intervals, some of the previously lost information regarding the
state of the impurity is temporarily recovered, thanks to the fact
that the dephasing coefficient $\gamma (t)$ becomes temporarily
negative.

{\itshape Dephasing dynamics in the limiting cases -} We assume
the total system to be initially at zero temperature, with the gas
prepared in its ground state, and the impurity in an equal
superposition state. In the superfluid regime, $U\ll J$, the
kinetic energy of the atoms in the lattice is much larger than the
on-site interaction, and hence the latter can be treated as a weak
perturbation. The gas is a 1D quasi-condensate whose majority of
atoms occupy the $k=0$ momentum state. It can be described using
Bogoliubov mean-field theory \cite{bhbog}, which allows us to recast
the local density operator as follows
 \begin{equation} \hat n_{0}=n_{0}+\sum_k \frac
{\beta_k}{N_{s}} \left(\hat b_k^\dagger+\hat b_k \right),
\label{bogintham}
\end{equation} in which $n_{0}$ is the condensate density,
$\beta_{k}=\sqrt{n_{0}N_{s}}\left(u_{k}+v_{k}\right)$ is the
spectral density expressed in terms of the Bogoliubov coefficients
$u_{k}, v_{k}$ \cite{bhbog}, and $\hat{b}^{\dagger}_{k}$ create
the Bogoliubov quasi-particle excitations with momentum $k$ (with $N_s$ being the
number of lattice sites).
In the strongly interacting regime $U\gg J$, we use the doublon-holon approximate description instead, derived in
\cite{huber2007,barmettler2012} and successfully employed to study
out-of-equilibrium dynamics \cite{daley}. We assume that, if the
gas filling factor is $\bar{n}$, defining the ground state
$\ket{\textrm{GS}}\propto\otimes_{j}\ket{\bar{n}}_{j}$ in the
limit $U/J \rightarrow \infty$, then only the local states
$\ket{\bar{n}}$, $\ket{\bar{n}+1}$ and  $\ket{\bar{n}-1}$ play a
major role in the dynamics.  Truncating the local basis to these
three states \cite{barmettler2012}, we obtain an effective
description of the system in terms of fermionic excitations,
giving
\begin{equation}
\begin{aligned}
\hat{n}_{0}=\bar n+\frac{1}{N_s^2}\sum_{k,q}\cos\left(\frac{\theta_{k}-\theta_{q}}{2}\right)(\hat{d}^{\dagger}_{k}\hat{d}_{q}-\hat{h}^{\dagger}_{k}\hat{h}_{q})+\\
i\sin\left(\frac{\theta_{k}-\theta_{q}}{2}\right)\left(\hat{d}^{\dagger}_{k}\hat{h}^{\dagger}_{-q}-\hat{h}_{-q}\hat{d}_{k}\right),
\label{mottinter}
\end{aligned}
\end{equation}
where $\hat{d}^{\dagger}_{k}$ ($\hat{d}_{k}$) and
$\hat{h}^{\dagger}_{k}$ ($\hat{h}_{k}$) are fermionic operators
creating (destroying) doublon and holon excitations, respectively
\cite {mottdetails}. This approach breaks down for $U/J < 4 (\bar
n +1)$, therefore choosing $\bar n=1$ restricts the range
of validity to $U/J  \ge 8$.

In the two limiting cases, analytic expressions for the dephasing
rate can be derived by plugging the density fluctuations of
\eqref{bogintham} and \eqref{mottinter} into the expression for
$\gamma (t)$. Including the constant density component in the
unitary evolution, and taking the average over the respective
vacuum states, we obtain:
\begin {align}
\gamma_{SF} (t) &=U_e^2 \sum_{k}\frac{|\beta_{k}|^{2}}{N_{s}^{2}}\left(\frac{\sin \omega_{k}t}{\omega_{k}^{}}\right), \label{gammasuno} \\
\gamma_{M} (t) &= U_e^2 \sum_{k,q}\sin^{2}\left(\frac{\theta_{k}-\theta_{q}}{2}\right)\frac{\sin(\omega^{d}_{k}+\omega^{h}_{q})t}{N^2_s(\omega^{d}_{k}+\omega^{h}_{q})^{}},
\label{gammas}
\end{align}
where $\omega_k$ and $\omega^{d/h}_k$ are the energies of the
superfluid and doublon/holon excitations, respectively \cite
{spectra}. As we discuss below, while $\gamma_{SF}$ gives rise to
a purely Markovian impurity dynamics (due to delocalised
excitations), $\gamma_M$  results in a generally non-zero
information backflow (due to the doublon-holon pairs, generating
localised particle density fluctuations).
\begin{figure}[!t]
\centering
\includegraphics[width = 1 \columnwidth, unit=1pt]{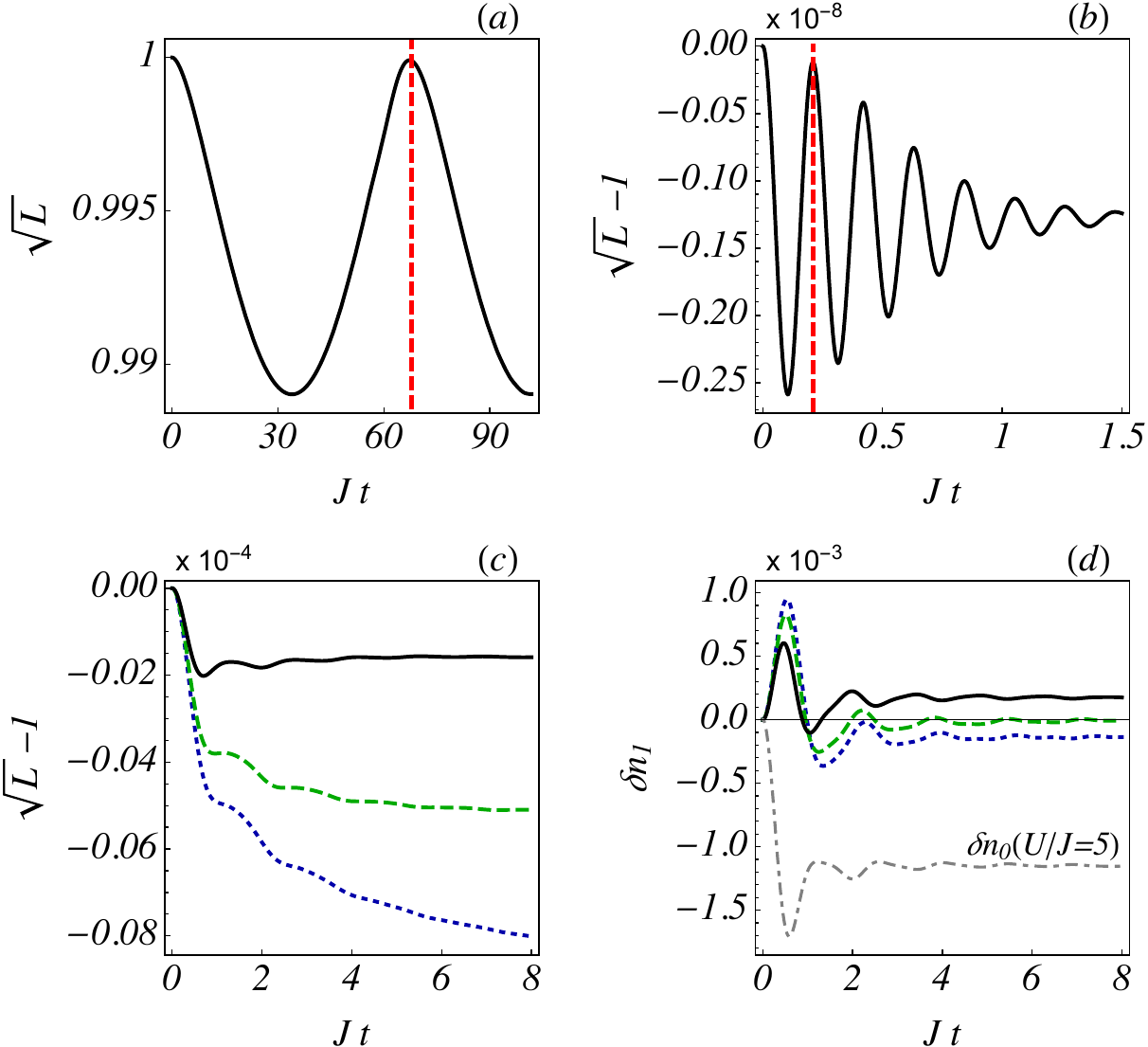}
\caption{(Color online). $(a)$: Loshmidt echo in the superfluid
phase calculated using Eq. \eqref {gammasuno} at $U/J=1$, the vertical line
is the recurrence time $\tau =\frac {N_S}{c_s} \simeq  \frac {2
\pi} {\omega_{k= {2 \pi}/{N_s}}}$, due to finite size effects.
$(b)$: Loshmidt echo deep in the insulating phase at $U/J=30$, the
vertical line is $\tau = \frac{2 \pi} {U} $. $(c)$: Loshmidt echo
in the intermediate regime for $U/J=3.5,4,5$ in dotted blue,
dashed green an solid black, respectively. $(d)$: density
fluctuations in the nearest neighbor of the quenched site (dubbed
as $1$) for $U/J=3.5,4,5$ in dotted blue, dashed green an solid
black, respectively. The dot-dashed gray line gives the density
fluctuation in the perturbed site for $U/J=5$. In all the panels
$U_e/J=0.01$ and $N_s=96$. Data displayed in $(b)$, $(c)$ and
$(d)$ were obtained trough particle-conserving ($\bar{n}=1$)
t-DMRG calculations.} \label{Plot1}
\end{figure}

{\itshape Results -} As seen from Eq. \eqref {nmopt},
non-Markovian effects occur if the Loschmidt echo does not decay
monotonically (requiring a temporary negative $\gamma(t)$). In the
superfluid regime, $L(t)$ displays revivals at long times only,
due to finite size effects, see panel $(a)$ of Fig. \ref {Plot1}.
The memory-inducing mechanism, in this case, is the following:
after the impurity is embedded in the Bose lattice at site $0$, representing the center of the lattice,
the repulsion generates a small depletion of the quenched site,
and a fraction of the initial particles moves towards the
neighbouring sites. In the superfluid regime, a density wave is
generated, which travels freely throughout the lattice. Revivals
in the density of the quenched site (and, as a consequence, in the
Loschmidt echo) are due to this wave reaching the boundary -
because of the finite size - and bouncing back towards its source.
In the Bogoliubov mean field approach, delocalised phononic
excitation have the effective speed of sound $c_s = \sqrt {2 J U
n_0}$, so that we can predict the revival time to be $N_s/c_s$.

In the Mott phase, the behaviour is qualitatively different.  At
strong interactions, the energy of a pair of two distinct types of
quasiparticle with opposite momenta is $2 \Omega (k) \simeq U -2 J
(2 \bar n +1) \cos k$ \cite {barmettler2012, spectra}. In this regime, the maximal relative
velocity of a pair is found at $|k| \simeq \frac {\pi} {2}$, where
the spectral density exhibits a maximum. As a result, the
decoherence function shows the first revival at $t\simeq \frac {2
\pi} {2 \Omega (\pi/2)}= \frac {2 \pi} {U}$, in agreement with the
numerics, see panel $(b)$ of Fig. \ref {Plot1}. Hence, the
non-monotonicity of the Loschmidt echo occurs for shorter and
shorter times with increasing the interaction strength. In the
limit $U/J\to\infty$, the gas approaches the hard-core boson
limit, with the exact eigenstates of the Hamiltonian in
Eq.~\eqref{bh} being given by local Fock states $\ket{n}_{j}$. In
this case, the interaction with the impurity
only causes a local energy shift, generating no dynamics and no
memory effects at all.

Around the critical point $(U/J)_c\approx3.4$
\cite{kuhn2,kollath,pino}, the description  gets more involved as
mean field approaches do not hold anymore. In analogy to the
limiting cases discussed above, we can start off considering the
effect of the impurity on the local particle density. As expected,
a depletion of the quenched site occurs; however, the local
density displays a few oscillations only, before reaching a
stationary value. In the Mott phase, the response of the gas after
 the embedding of the impurity can be witnessed by analysing the
density fluctuations at the neighbouring site $i=1$, $\delta{n}_{1}(t)=\langle\hat{n}_{1}(t)\rangle-\langle\hat{n}_{1}\rangle_{\textrm{GS}}$. For $U/J
\lesssim 4$, after a quick transient with a positive density
fluctuation, a negative stationary value is reached, see panel
$(d)$ of Fig. \ref {Plot1}. After $U/J \simeq 4$ the extra
particles moving away from site $0$ stay almost localized, as they
oscillate back and forth, accumulating (on average) on site $1$,
i.e. generating a positive stationary value for the density
fluctuation.

To understand this behavior, we can compare it with the results of
Ref. \cite {Ronzheimer2013}, where the expansion velocity $v_{c}$
of an initially held cloud of atoms in a 1D Bose-Hubbard lattice
was thoroughly investigated and a minimum was found at
$(U/J)_{v_{c}}\simeq 4$, consistently with other experimental observations \cite{Trotzky2012}. In that scenario, high occupancy states
begin to form at small $U/J$ (but still in the Mott phase),
resulting in a reduction of the expansion velocity of the cloud.
In our setup, where the lattice is not switched off, the excess of
particles generated by the impurity ends up trapped in the
vicinity of the quenched site for $U/J \gtrsim 4$. In the
intermediate regime (between the critical point and $U/J \simeq
4$), instead, they are able to propagate even though the
environment has entered the insulating phase. This behaviour can
be related to the emergence of structures in the Loschmidt echo.
Proper oscillations are found for $U \gtrsim 4$, with $L(t)$
displaying a series of maxima and minima that signal a
non-Markovian dynamics, ultimately due to the localized character
of the density fluctuations, see panel (c) in Fig. \ref {Plot1}.
On the other hand, for $U \lesssim 4$ the Loshmidt echo displays a
series of inflection points, not giving rise to any true memory
effect. Finally, all these structures disappear in the superfluid
regime. \par
\begin{figure}[!t]
\centering
\includegraphics[width =0.85 \columnwidth, unit=1pt]{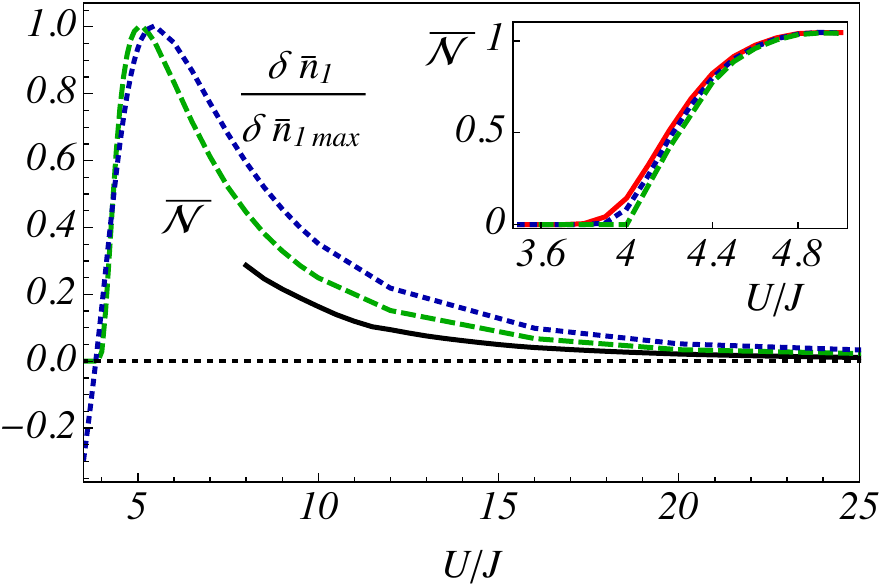}
\caption{(color online):  Comparison between numerical (dashed
green, $N_{s}=96$) and analytical (solid black, $N_{S}\to\infty$)
normalised non-Markovianity measure
$\overline {\mathcal{N}} =\mathcal{N}/\mathcal{N}_\textrm{max}$ and average excess particle
number at site $1$,
$\delta\bar{n}_{1}/\delta\bar{n}_{1\textrm{max}}$ (dotted blue,
$N_{s}=96$), where the time average is performed over $JT=2 \pi $.
In the inset, $\overline {\mathcal{N}} $ for different lattice sizes, $N_s=80$, $96$ and $128$ in solid red, dotted blue and dashed green respectively.} \label{Plot2}
\end{figure}

The memory quantifier displays the counterpart of these different
behaviours of the local density. In Fig. \ref{Plot2}, we display
the normalized non-markovianity measure,
$\overline {\mathcal{N}}=\mathcal{N}/\mathcal{N}_{\textrm{max}}$, comparing a numerical
calculation 
with the analytic results obtained with the doublon-holon model in
the thermodynamic limit \cite{note}. The plot summarises our
findings, showing that the two limiting phases of the lattice gas
are characterised by qualitatively different dynamics: fully
Markovian when the environment is in the superfluid phase and
non-Markovian when the environment is deep in the Mott insulator
phase. The Markovian-to-non-Markovian transition occurs at
$(U/J)_{\mathcal N} \simeq 4$, different from but still quite
close to the critical point. Notice, in particular, the good
agreement between the numerical and the approximate analytic
results, discrepancies being due to the perturbative nature of the
master equation and to the truncation of the Hilbert space adopted
in the analytic approach. 
Increasing the ratio $U/J$, the two calculation methods agree in
giving an amount of information backflow that decays
asymptotically to zero. The onset of non-Markovian dynamics is
then compared with the time averaged occupancy fluctuation of the
impurity's adjacent site
$\delta\bar{n}_{1}=1/T\int_{0}^{T}\langle\hat{n}_{1}(t)\rangle-\langle\hat{n}_{1}\rangle_{\textrm{GS}}$.
This quantity is negative for values of $U/J$ at which the gas has
a larger mobility, resulting in quasi-particles being pushed away
from the neighbourhood of the impurity and spreading without any
backflow, hence leading to Markovian dynamics. As soon as
$\delta\bar{n}_{1}$ becomes positive, instead, extra particles
moving away from site $0$ oscillate back and forth, accumulating
(on average) on site $1$, and giving rise to the observed memory
effects. This is true up to a maximum after which the gas
approaches the hard-core boson limit, where even these
oscillations are strongly suppressed, resulting in a decay of
$\delta\bar{n}_{1}$ back to its ground-state value, and in a
tendency of the dynamics to become memory-less again.
In this system, therefore, we can understand the quasi-particles
as the physical information carriers, their dynamics being
uniquely connected to information backflow and non-Markovian
memory effects.

{\itshape Conclusions and outlook -} We have investigated the open
system dynamics induced by a Bose-Hubbard lattice when it plays
the role of environment for an (open) impurity system. We have
derived the generalized master equation for the open system
dynamics and linked the dephasing coefficient to the
density-density correlations of the cold atom environment. Using
both analytical and numerical techniques we have studied the
non-equilibrium dynamics of the total system and we have shown
that the ratio $U/J$ can be thought of as a  control parameter
allowing to manipulate the nature of the information carriers and
thus ruling the Markovian to non-Markovian crossover. In this
sense, the Bose-Hubbard lattice can act as a controllable
engineered environment, allowing for the simulation of open
quantum system dynamics in both Markovian and non-Markovian
regimes. We have shown that in this system the presence or absence
of memory effects is directly linked to the nature of the
excitations induced in the environmental ground state and showed
how the onset of memory effects  signals the trapping of
higher occupancies in the vicinity of the impurity.

{\itshape Acknowledgements-}  The authors acknowledge support from the Horizon 2020 EU collaborative
projects QuProCS (Grant Agreement No. 641277). D. J. acknowledges EPSRC support by projects EP/P009565/1 and EP/K038311/1. The authors also acknowledge the use of the University of Oxford
Advanced Research Computing (ARC) facility in carrying out this work. http://dx.doi.org/10.5281/zenodo.22558. This research is partially funded by the European Research Council under the European Union’s Seventh Framework Programme (FP7/2007-2013)/ERC Grant Agreement no. 319286 Q-MAC. This work was also supported by the EPSRC National Quantum Technology Hub in Networked Quantum Information Processing (NQIT) EP/M013243/1. J. J. M. A. acknowledges financial support from Facultad de Ciencias at UniAndes-2015 project ”Quantum
control of non-equilibrium hybrid systems-Part II”.


\begin{thebibliography} {99}
\bibitem{note}
Discrepancies between the analytical models and numerical
computation, for values of $U/J$ allowing for a direct comparison,
arise from two key differences. The first is the dimension of the
Hilbert space of the gas; while in the tDMRG approach local Fock
states up to $n=5$ are allowed, the fermionic model introduced
above for the strongly interacting gas only includes $n=0,1,2$.
Second, unlike the quasi-exact tDMRG method, the analytical
results rely on a perturbative approach in deriving the open
dynamics of the impurity. Despite such differences, a good
agreement between analytical and numerical results is found.
\bibitem{mottdetails}  The Bogoliubov angles are
$\theta_{k}=\arctan\left[-2i\Delta_{k}/(E^{d}_{k}+E^{h}_{k})\right]$,
with $E^{d}_{k}=-2J(\overline{n}+1)\cos k+U/2,
E^{h}_{k}=-2J\overline{n}\cos k+U/2,
\Delta_{k}=2iJ\sqrt{\overline{n}(\overline{n}+1)}\sin k$.
\bibitem{spectra}  The energies of the superfluid excitations are given by
$\omega_k =\sqrt {4 J U n_0 (1- \cos k)+4 J^2 (1-\cos k)^2} $. The energies of the quasi-particles in the doublon/holon description are
$\omega^{d/h}_{k}=\mp J\cos k+\Omega_{k}$, with
$\Omega_{k}=\frac{1}{2}\sqrt{(E^{d}_{k}+E^{h}_{k})^{2}+4|\Delta_{k}|^{2}}$.
\end{thebibliography}
\end{document}